\documentclass[runningheads]{llncs}
\usepackage[T1]{fontenc}
\usepackage{graphicx,verbatim}

\usepackage{amssymb, amsmath, bm, latexsym,comment}
\usepackage{multirow}
\usepackage{xcolor}
\usepackage{subfigure}
\usepackage{indentfirst}
\usepackage{setspace}
\usepackage{verbatim}
\usepackage{array}
\usepackage{arydshln}
\usepackage[misc]{ifsym} 

\providecommand{\citep}[1]{\cite{#1}}

\begin{document}

\title{Personalized 3D Myocardial Infarct Geometry Reconstruction from Cine MRI with Explicit Cardiac Motion Modeling}
\titlerunning{3D Myocardial Infarct Geometry Reconstruction}


\author{Yilin Lyu \inst{1} \and 
Fan Yang \inst{1} \and 
Xiaoyue Liu \inst{1} \and 
Zichen Jiang \inst{2} \and 
Joshua Dillon \inst{3} \and 
Debbie Zhao \inst{3} \and 
Martyn Nash \inst{3} \and 
Charlene Mauger \inst{4} \and 
Alistair Young \inst{4} \and 
Ching-Hui Sia \inst{5,6} \and 
Mark YY Chan \inst{5,6} \and 
Lei Li \inst{1}${^{(\textrm{\Letter})}}$ } 
\authorrunning{Lyu et al.}
\institute{Department of Biomedical Engineering, National University of Singapore, Singapore \and
Department of Computer Science, National University of Singapore, Singapore \and
Auckland Bioengineering Institute, University of Auckland, Auckland, New Zealand \and
School of Biomedical Engineering and Imaging Sciences, King’s College London, UK  \and
Department of Medicine, National University of Singapore, Singapore \and
Department of Cardiology, National University Heart Centre Singapore, Singapore \\
\email{lei.li@nus.edu.sg}}


\maketitle  
\begin{abstract}

Accurate representation of myocardial infarct geometry is crucial for patient-specific cardiac modeling in MI patients.
While Late gadolinium enhancement (LGE) MRI is the clinical gold standard for infarct detection, it requires contrast agents, introducing side effects and patient discomfort.
Moreover, infarct reconstruction from LGE often relies on sparsely sampled 2D slices, limiting spatial resolution and accuracy.
In this work, we propose a novel framework for automatically reconstructing high-fidelity 3D myocardial infarct geometry from 2D clinically standard cine MRI, eliminating the need for contrast agents. 
Specifically, we first reconstruct the 4D biventricular mesh from multi-view cine MRIs via an automatic deep shape fitting model, biv-me.
Then, we design a infarction reconstruction model, CMotion2Infarct-Net, to explicitly utilize the motion patterns within this dynamic geometry to localize infarct regions.
Evaluated on 205 cine MRI scans from 126 MI patients, our method shows reasonable agreement with manual delineation.
This study demonstrates the feasibility of contrast-free, cardiac motion-driven 3D infarct reconstruction, paving the way for efficient digital twin of MI.


\end{abstract}

\section{Introduction}

Myocardial infarction (MI) remains a major cause of mortality and disability worldwide \cite{journal/Lancet/reed2017}. 
Structural and electrophysiological remodeling in infarcted regions plays a key role in post-MI complications, including arrhythmias \cite{journal/FiP/mendonca2018}. 
Recently, computational modeling of patient-specific hearts has emerged as a promising non-invasive tool for guiding personalized treatment \cite{journal/EHJ/corral2020,journal/EP/wang2021}. 
Accurately representing myocardial remodeling in ischemic cardiomyopathy requires integrating patient-specific infarct geometry into these models \cite{journal/TMI/li2024}.
Among clinical imaging techniques for infarct characterization, late-gadolinium enhanced magnetic resonance image (LGE MRI) is the most widely used \cite{journal/MedIA/li2023}. 
While effective, LGE MRI requires contrast agent injection, which may cause side effects, increase scanning time, and reduce patient comfort \cite{journal/SR/polacin2021}. 
In contrast, cine MRI, a standard clinical tool for visualizing cardiac anatomy and motion, offers non-invasive imaging of the heart without contrast agents. 
However, both LGE and cine MRI typically capture sparse, intersecting 2D planes, i.e., short-axis (SAX) and a few long-axis (LAX) slices, limiting spatial resolution and hindering the reconstruction of a detailed 3D heart model.

For 3D heart model reconstruction from 2D cardiac planes, many work employ two-stage model, i.e., image segmentation and 3D geometry reconstruction \cite{conf/ICCV/ye2023,conf/ISBI/biffi2019,conf/ICCV/yuan2023,journal/MedIA/laumer2023,journal/arxiv/chen2024}.
However, all these work only can reconstruct the 3D (or 3D + t) geometry model, where the infarction area is not identified. 
The computational modeling needs 3D infarct geometry, which is still coarsely estimated based on scar interpolation from LGE MRI \cite{journal/TMI/ukwatta2015}.
Recently, several studies have explored scar analysis using contrast-free imaging as a more cost-effective alternative \cite{journal/MedIA/xu2020a,journal/MedIA/xu2018,journal/Radiology/zhang2019}.
One widely adopted approach leveraged generative models to synthesize LGE-style images, enabling LGE-based analyses without the need for contrast agents. 
For example, Xu et al. \cite{journal/MedIA/xu2020b} introduced sequential causal generative models that integrated synthesis and scar segmentation within an adversarial learning framework.
However, these methods are highly dependent on the quality of the synthesized LGE images.  
An alternative strategy is to analyze cine MRI directly, utilizing its temporal information to detect abnormalities \cite{journal/MedIA/xu2018,journal/Radiology/zhang2019,conf/ISBI/yang2025}. 
However, all these works only capture the cardiac motion information in the 2D single view. 
Moreover, they only implicitly extract the cardiac motion information such as optical flow-based methods, which primarily capture motion between adjacent frames.

In this study, we develop a 3D infarct geometry reconstruction model that leverages cardiac motion features extracted from multi-view cine MRIs. 
The proposed framework explicitly integrates cardiac morphology and motion dynamics to establish a relationship between abnormal myocardial motion and infarcted regions. 
This is accomplished by introducing a 4D cardiac mesh, derived from cine MRI, as the input to a cardiac motion mapping model (CMotion2Infarct-Net) for infarction localization. 
Furthermore, by leveraging the spatial correspondence between cine and LGE MRI, we project the 2D scar information from LGE MRI onto the reconstructed cardiac mesh, providing supervision for CMotion2Infarct-Net training. 
To the best of our knowledge, this is the first study to directly reconstruct a 3D myocardial infarct model from cine MRI.

\begin{figure*}[t]\center
 \includegraphics[width=0.99\textwidth]{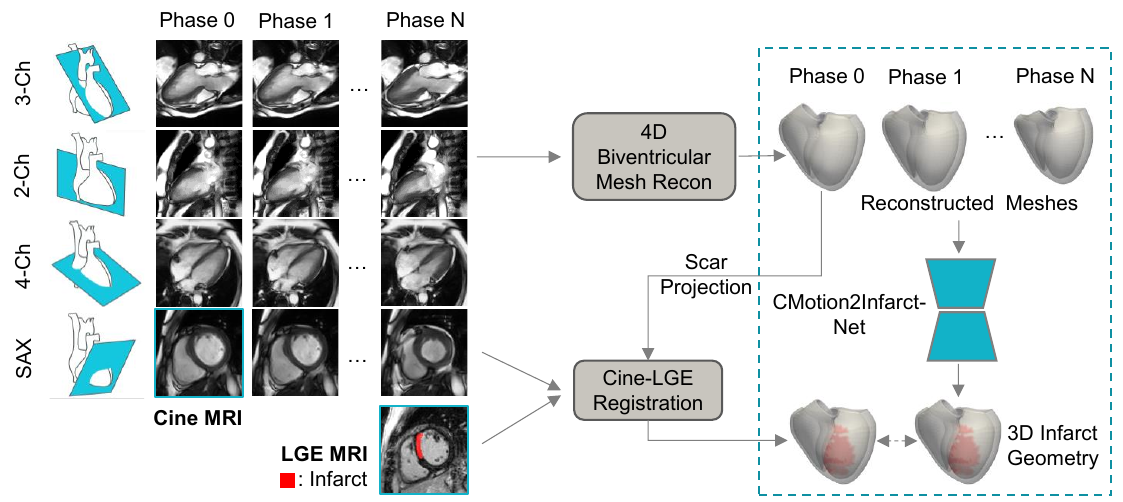}\\[-2ex]
   \caption{Illustration of the multi-view cine-MRI based 3D infarct reconstruction framework. Note that the figure only presents a single short-axis (SAX) slice as an example, even though a stack of SAX view are employed here. LGE MRI is registered to the SAX end-diastolic (ED) phase, i.e., phase 0.}
\label{fig:method:framework}
\end{figure*}

\section{Methodology}

Fig.~\ref{fig:method:framework} provides an overview of the proposed 3D infarct
geometry reconstruction model, consisting of 4D biventricular mesh reconstruction module, cine-LGE registration module, and cardiac motion mapping to infarct model.
The 4D mesh reconstruction module integrates four different views of cine MRI together and fits them to a mesh template for cardiac reconstruction (Sec.~\ref{method:4dheart}). 
To train the CMotion2Infarct-Net, we generate the 3D Ground Truth (GT) infarct model as supervision based on cine and LGE registration and 3D scar projection (Sec.~\ref{method:registration}). 
Finally, Sec.~\ref{method:inference} presents the details of the reconstruction model for the personalized inference of 3D infarct model.

\subsection{4D Biventricular Model Reconstruction from Cine MRI} \label{method:4dheart}

We adopt biv-me \cite{bivme2025}, an open-source and fully automated reconstruction pipeline, to infer 4D biventricular meshes from multi-view cine MRIs.
It consists of three stages: view selection, segmentation, and cardiac geometric fitting.
ResNet50 is first employed to identify and classify useful views within the cine MRI sequences.
Next, the nnU-Net is used to segment the biventricular region, i.e., left ventricle (LV) cavity, right ventricle (RV) cavity, LV myocardium, and extract corresponding 2D contours from the selected views. 
Finally, these sparse contour sets are merged together based on their world coordinate and used to reconstruct biventricular meshes for each time frame through an iterative diffeomorphic registration algorithm. 
This is achieved by decomposing the deformations into two steps to ensure a bijective transformation. 
Specifically, a multi-class surface template mesh is first aligned to each contour set using an implicit linear least squares fit.  
The successive least-square fits can accelerate convergence and improve initialization at a lower computational cost. 
To preserve topology, the displacement of the coarse mesh is constrained within each iteration, ensuring that the Jacobian determinant remains positive.
The explicit diffeomorphic fit further refines the alignment, ensuring a structured, point-correspondent mesh representation across the cardiac cycle.

\subsection{Registration of Cine and LGE MRI for 3D Scar Projection} \label{method:registration}

To generate a 3D representation of the infarct region for supervision, we leverage manual scar segmentation from LGE MRI and project the identified scars onto a 3D biventricular surface mesh reconstructed from cine MRI.  
Due to differences in spatial resolution, field of view, and respiratory motion, cine and LGE MRI are often misaligned.
To address this, we employ a multivariate mixture model-based registration framework \cite{journal/TPAMI/zhuang2018} to align cine and LGE MRI.
The registration process involves identifying corresponding slices along the Z-axis, followed by in-plane rigid and non-rigid transformations. 
Once LGE MRI is spatially aligned with cine MRI, the LGE-derived annotations can be accurately mapped onto end-diastolic (ED) phase cine images.  
Nonetheless, the transformed infarct regions remain sparse due to the inherent slice thickness and limited coverage of 2D MRI acquisitions. 
Accordingly, we employ Gaussian sampling to generate denser scar distribution. 
Specifically, for each scar voxel identified in cine MRI, additional points are synthesized along the Z-axis by sampling from a normal distribution  
\(\mathcal{N}(\mu, \sigma^2)\), where \(\mu\) is the original Z-coordinate and \(\sigma = 3\) mm.  
These augmented scar points are then mapped onto the 5 nearest vertices of the 3D heart surface mesh using a KDTree-based nearest-neighbor search. 
Note that for simplification in this study we project all scars onto LV endocardium, as used in Codreanu et al. \cite{journal/JMCC/codreanu2008}.
Subsequently, the corresponding mesh vertices are labeled as scarred regions, ensuring a more spatially coherent and anatomically realistic infarct representation on the 3D surface mesh.


\begin{figure*}[t]\center
 \includegraphics[width=0.99\textwidth]{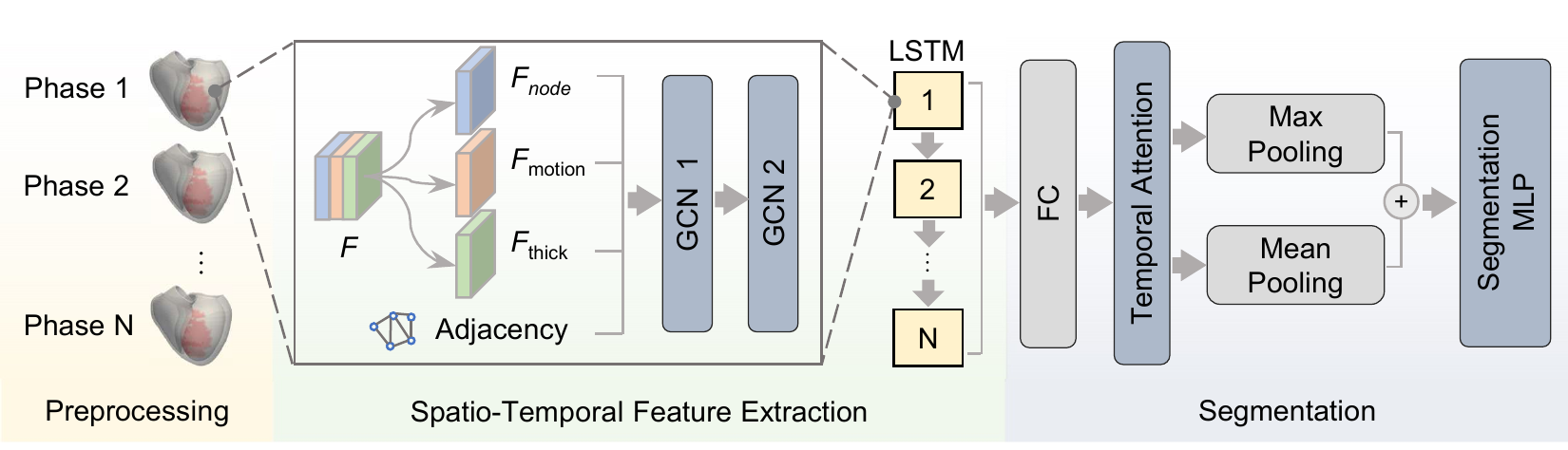}\\[-2ex]
   \caption{The architecture of CMotion2Infarct-Net. N denotes the index of the phrase, \( F\) represents the features after preprocessing,  \( F_{\text{node}} \), \( F_{\text{motion}} \), \( F_{\text{thick}} \) correspond to the features representing position, inter-phrase dynamics, and wall thickness, respectively.}
\label{fig:method:model architecture}
\end{figure*}

\subsection{Explicit Cardiac Motion based Infarct Reconstruction} \label{method:inference}

Given a sequence of 4D biventricular surface meshes, \(\{M_t\}_{t=1}^{N}\), their spatial and motion characteristics can be captured using CMotion2Infarct-Net. 
CMotion2Infarct-Net comprises a preprocessing module, a spatio-temporal feature extraction module, and an attention based segmentation module.
The module takes the biventricular model as input, while the network focus is on the left ventricular myocardium.
So, we extracted the endocardial mesh of LV and combined it with the corresponding epicardial point cloud as hybrid input for the next step.
It is well recognized that an important feature of MI is the abnormal motion \cite{journal/CAR/Feldmann2019,journal/CHFR/thune2006,journal/Echo/alam1999,journal/IHJ/cerisano2001}.
Therefore, to further enhance the extraction of local features, we introduced first-order inter-phase differences as the local motion feature $F_{\text{motion}}$.
Moreover, there are relationships between infarct regions and wall thickening \cite{journal/CBM/yousefibanaem2017,journal/JCMR/newosielski2009}.
Accordingly, we also introduced thickness $F_{\text{thick}}$ as the initial input of CMotion2Infarct-Net.
The spatio-temporal feature extraction module first utilizes graph neural network (GNN) to extract structural features. 
After that, a two-layer long short-term memory (LSTM) network is applied to capture temporal dependencies across all phases, enabling point-wise feature extraction that integrates both spatial and temporal information.
The segmentation head consists of a fully connected (FC) layer, a temporal attention layer, two spatial pooling operations (max pooling and mean pooling), and a MLP.
Here, we design a transformer with 4 attention head to capture long dependencies and complex temporal interactions.
Then, we utilize max pooling and mean pooling to extract global spatial features and concatenate for the next step.
And finally, a two-layer segmentation MLP maps the latent features to the final segmentation output.
CMotion2Infarct-Net is optimized by minimizing the regularized mesh segmentation loss, formulated as:  
\begin{equation}
    \mathcal{L}(M_{infarct}, \hat{M}_{infarct}) = \mathcal{L}_{BCEweighted} + \lambda_{Tversky} \mathcal{L}_{Tversky}(\alpha, \beta)
\end{equation}
where \(M_{infarct}\) and \(\hat{M}_{infarct}\) represent the predicted and GT infarct, respectively.
Since the infarct region typically occupies smaller area compared to the normal LV, $\mathcal{L}_{BCE}$, specifically a weighted binary cross entropy (BCE) loss is employed as the primary loss.
$\mathcal{L}_{Tversky}$ represents the Tversky Loss, and $\alpha$, $\beta$ are weighting parameters to balance false positives and false negatives, respectively.
$\lambda$ are balancing parameters.

\section{Experiments and Results}

\subsection{Materials}

\subsubsection{Data Acquisition.}

We collected 205 paired LGE and multi-view cine MRI scans from 126 post-MI patients across multiple centers.
Specifically, a stack of SAX balanced steady-state free precession (bSSFP) cine MRI and three LAX cine sequences (2-, 3-, and 4-chamber views) have been acquired, as shown in Fig. \ref{fig:method:framework}. 
The SAX cine sequences consist of 8 to 17 slices across 25 frames.
The dataset was randomly divided into 150 for training, 10 for validation, and 45 for test.
To prevent data leakage, each patient's data is included in only one dataset.
As a result, we use data from 93 patients for training, 6 for validation and 27 for test, respectively.

\subsubsection{Implementation.} 

The framework was implemented in Pytorch, running on a workstation equipped with an AMD EPYC 7K62 Processor and a NVIDIA GeForce RTX 4090 GPU. 
We use the Adam optimizer to update the network parameters via stochastic gradient decent ($\text{weight decay} = 1 \times 10^{-4}$). 
The initial learning rate is set to $1 \times 10^{-4}$ and multiplied by 0.7 every approximately 800 iterations.
The parameters in Sec.~\ref{method:inference} are set as follows: \(\lambda_{Tversky} = 1\), and $\alpha$=0.3, $\beta$=0.7 to penalize false negatives more.
The biv-me took about 9 min for each subject.
CMotion2Infarct-Net was trained in 2 hours (600 epochs), with an inference time of 5 seconds per case.

\subsubsection{Gold Standard and Evaluation.}

To evaluate the reconstruction accuracy of 4D mesh, we manually segmented the biventricular area (LV, RV, and LV Myo) using ITK-SNAP on the cine data at the ED phase. 
We also calculated time-resolved chamber volumes for each reconstructed 4D heart.
For 3D infarct evaluation, LGE MRIs were manually segmented by a trained student and verified by an expert.
Ground-truth infarct regions on cine MRI were obtained by registering LGE segmentation to ED-phase cine images (Sec. \ref{method:registration}).
The generated 3D infarct geometry is used as the GT in this study.
We then employed Dice score, Recall, ASD and Generalized Dice ($\textit{G}$ Dice) to assess the infarct region overlap and alignment between the predicted infarct geometry and the GT. 

\subsection{Results}

\subsubsection{Accuracy of 4D Biventricular Reconstruction.} 

\begin{figure*}[t]\center
    \subfigure[] {\includegraphics[width=0.1\textwidth]{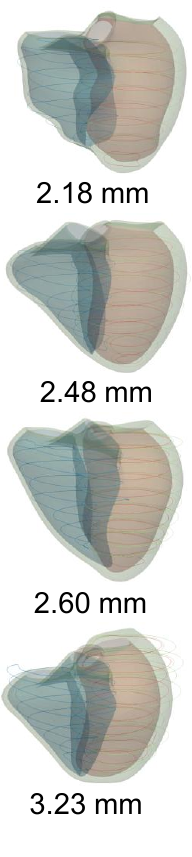}}
    \subfigure[] 
    {\includegraphics[width=0.86\textwidth]{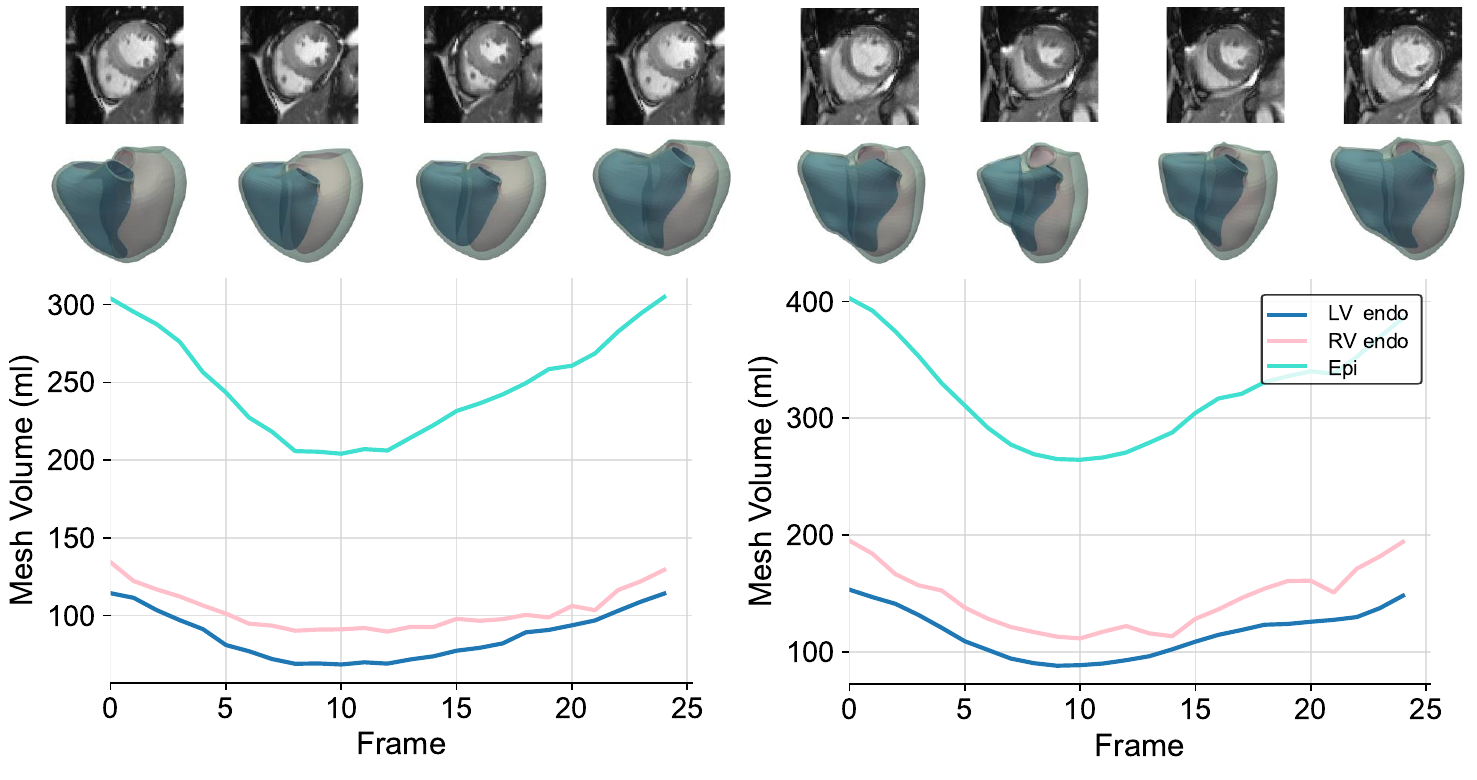}}
  \caption
    {(a) 3D visualization of the overlap between the sparse contours and reconstructed heart; (b) Illustration of cine MRI in short-axis view and the predicted 4D biventricular mesh with corresponding volume change over time.}
\label{fig:result:mesh}
\end{figure*}


The average ASD between manually segmented contours and the 4D meshes reconstructed by biv-me is $2.55 \pm 0.452$ mm across 49 randomly selected subjects.  
Fig. \ref{fig:result:mesh} (a) presents the overlap visualization of the biv-me predicted meshes and GT for four representative subjects. 
The predicted meshes are closely aligned with the sparse contours.
To further assess motion accuracy, we analyzed volume change curves of the reconstructed 4D meshes. 
Fig. \ref{fig:result:mesh} (b) illustrates two examples of the predicted heart shapes alongside the corresponding chamber volume variations over time. 
The results indicate that biv-me effectively captures the systolic and diastolic phases observed in cine MRI, highlighting its ability to preserve physiological motion dynamics.  
The alignment in shape and motion indicates accurate extraction of physiological deformation.

\begin{table*}[t]
\centering
\caption{Summary of the quantitative evaluation results of 3D infarct reconstruction.}
\label{tb:result:ASD}
\footnotesize 
\begin{tabular}{c|c c c c c}
\hline
Method & Dice & Recall & ASD (mm) & $\textit{G}$ Dice\\
\hline
Inter-observer variation & $0.798 \pm 0.087$ & $0.770 \pm 0.125$ & $1.158 \pm 0.550$ & $0.824 \pm 0.079$ \\
\hline
CMotion2Infarct-Net & $\textbf{0.652} \pm 0.174$ & $0.805 \pm 0.179$ & $\textbf{3.135} \pm 2.911$ & $\textbf{0.686} \pm 0.162$ \\
w/o\! Temporal Attention & $0.611 \pm 0.181$ & $0.798 \pm 0.175$ & $3.784 \pm 2.999$ & $0.643 \pm 0.176$ & \\
w/o\! \( F_{\text{motion}} \) & $0.519 \pm 0.159$ & $0.764 \pm 0.128$ & $5.554 \pm 3.158$ & $0.545 \pm 0.159$ & \\
w/o\! \( F_{\text{thick}} \) & $0.608 \pm 0.178$ & $\textbf{0.806} \pm 0.168$ & $3.883 \pm 3.075$ & $0.638 \pm 0.171$ & \\
\hline
\end{tabular}
\end{table*}

\begin{figure*}[t]\center
 \includegraphics[width=0.99\textwidth]{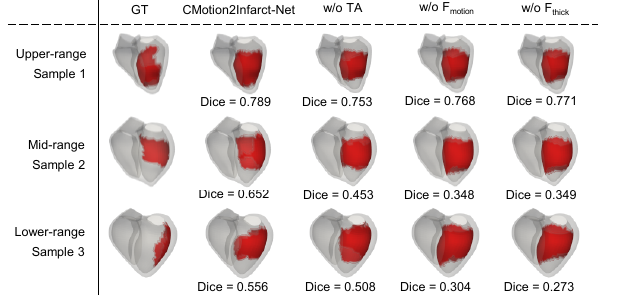}\\[-2ex]
   \caption{Comparison of prediction under different Dice scores. TA: Temporal Attention.}
\label{fig:results:visualization}
\end{figure*}

\subsubsection{Accuracy of 3D Infarct Reconstruction.} 

Similar to the Tversky loss in Sec.~\ref{method:inference}, we prioritized the identification of the positive class (i.e., infarct regions), so we emphasized Recall as a key metric.
Given that the infarct regions account for only $\sim 8.3\%$ of the total LV nodes, we incorporated $\textit{G}$ Dice score to better evaluate the performance under class imbalance \cite{conf/DLMIAS/sudre2017}.
We evaluated on the test set, and the results are presented in Table \ref{tb:result:ASD}.
CMotion2Infarct-Net achieved a reasonable Recall score of 0.805, indicating strong coverage of scar regions.
Although the average Dice score is 0.652, which reflected moderate overlap, the average $\textit{G}$ Dice score increases to 0.686, highlighting improved performance in identifying small infarct regions across different cases.
In addition, ASD suggested that the predicted boundaries are geometrically close to the GT, further validating the spatial precision.
Fig. \ref{fig:results:visualization} illustrates qualitative comparisons between predicted infarct regions and GT across different performance levels.
We selected representative test samples from the approximate top, middle, and bottom quartiles based on Dice scores for visualization, with infarct regions highlighted in red.
In worse cases, CMotion2Infarct-Net tends to over-predict infarct regions, resulting in some false positives or spatial over-coverage.
In contrast, high-performing cases exhibit strong agreement with the ground truth, while median cases show reasonable agreement with slight deviations.

Notably, one potential limitation of supervised approach is that the performance is affected by human expertise.
For instance, although most scars should have clearly defined boundaries, some ambiguous regions remain, especially scars near the atria or apex where image quality is worse, which may lead to inter-observer variation.
To further evaluate the reliability and stability, we invited an another expert to manually segment 15 cases.
A comparison with model predictions is shown in Table \ref{tb:result:ASD}.
The inter-observer Dice score is approximately 0.8, and $\textit{G}$ Dice score has even reached 0.824.
Our goal is to achieve accuracy comparable to human experts.
Despite some differences, preliminary results indicate that CMotion2Infarct-Net has achieved reasonable accuracy.

\subsubsection{Ablation study.} 

To evaluate the contributions of key modules and input features to model performance, we designed ablation experiments.
The experimental results are shown in Table \ref{tb:result:ASD}.
Regarding input features, we removed the motion feature $F_{\text{motion}}$ (which captures myocardial motion) and the thickness feature $F_{\text{thick}}$ (which reflects myocardial thickness), respectively to evaluate their independent contributions.
The experimental results showed that removing either feature led to performance degradation.
In particular, when the $F_{\text{motion}}$ was excluded, the all score dropped significantly, indicating that motion was crucial in identifying infarct regions.
Although $F_{\text{thick}}$ removal had little impact on Recall, it resulted in lower Dice (G Dice) and higher ASD, suggesting its complementary value.
Additionally, we removed the temporal attention module in CMotion2Infarct-Net to assess its effectiveness in modeling temporal dependencies.
Both the Dice and Recall scores decreased noticeably, confirming the importance of the Transformer in capturing temporal dynamics.
Some of the ablation results are also illustrated in Fig. \ref{fig:results:visualization} as a visual comparison with the original predictions.
The visual comparisons are consistent with the quantitative results: removing key modules leads to a decline in model performance, particularly in challenging cases (i.e., those around or below the median level).

\section{Conclusion}

In this work, we proposed an novel framework for automatic 3D infarct geometry reconstruction by combining multi-view cine MRIs. 
The proposed method fully leverages cardiac motion along the temporal dimension to enable structured representation of infarct regions without contrast agents, offering significant potential for clinical application.
The results have showed that the motion information can be accurately extracted from cine MRI and explicitly mapped to the region of infarction.
One major challenge is the large variability in infarct morphology, intensity, and spatial distribution.
In our future work, We will address these challenges and extend the surface-based infarct representation to a 3D volumetric (tetrahedral) model, enabling its direct use in personalized cardiac simulations and broader digital twin applications for post-infarction patient care.


\newpage
\bibliographystyle{splncs04}
\bibliography{A_refs}

\end{document}